\documentclass[twocolumn,trackchanges]{aastex631}

\usepackage{amsmath}

\begin{document}

\title{Rapid radio variability in the $z\sim7$ blazar VLASS J0410$-$0139: \\   
Indication of a hidden population of weak radio jets at Cosmic Dawn}

\author[0000-0003-4747-4484]{Silvia Belladitta}
\affiliation{Max-Planck-Institut f\"ur Astronomie, K\"onigstuhl 17, D-69117, Heidelberg, Germany}
\affiliation{INAF-Osservatorio di Astrofisica e Scienza dello Spazio, via Gobetti 93/3, I-40129, Bologna, Italy}

\author[0000-0002-2931-7824]{Eduardo Ba\~{n}ados}
\affiliation{Max-Planck-Institut f\"ur Astronomie, K\"onigstuhl 17, D-69117, Heidelberg, Germany}

\author[0009-0009-8274-441X]{Anniek J. Gloudemans}
\affiliation{NSF NOIRLab, Gemini Observatory, 670 N A'ohoku Place Hilo, HI 96720, USA}

\author[0000-0001-5648-9069]{Timothy W. Shimwell}
\affiliation{ASTRON, The Netherlands Institute for Radio Astronomy, Postbus 2, NL-7990 AA Dwingeloo, The Netherlands}
\affiliation{Leiden Observatory, Leiden University, PO Box 9513, 2300 RA Leiden, The Netherlands}

\author[0000-0003-3168-5922]{Emmanuel Momjian}
\affiliation{National Radio Astronomy Observatory, P.O. Box O, Socorro, NM 87801, USA}

\author[0000-0002-5880-2730]{Huib Intema}
\affiliation{Leiden Observatory, Leiden University, PO Box 9513, 2300 RA Leiden, The Netherlands}

\author[0000-0002-5941-5214]{Chiara Mazzucchelli}
\affiliation{Instituto de Estudios Astrof\'{\i}sicos, Facultad de Ingenier\'{\i}a y Ciencias, Universidad Diego Portales, Avenida Ej\'{e}rcito Libertador 441, Santiago, Chile}

\author[0000-0002-3528-7625]{Christian Fendt}
\affiliation{Max-Planck-Institut f\"ur Astronomie, K\"onigstuhl 17, D-69117, Heidelberg, Germany}

\author[0000-0001-5424-0059]{Bhargav Vaidya}
\affiliation{Department of Astronomy Astrophysics and Space Engineering, Indian Institute of Technology Indore, Khandwa Road, Simrol, Indore 453552, India}

\author[0000-0003-4793-7880]{Fabian Walter}
\affiliation{Max-Planck-Institut f\"ur Astronomie, K\"onigstuhl 17, D-69117, Heidelberg, Germany}



\begin{abstract}
Doppler boosting allows blazars to be detected out to high-$z$, making them promising probes of the intergalactic medium through the 21 cm forest. 
We report 0.144$-$11\,GHz observations of the most distant known blazar, VLASS\,J041009.05$-$013919.88 at z\,$\sim$\,7, obtained with the upgraded Giant Metrewave Radio Telescope (uGMRT), the LOw Frequency ARray (LOFAR) and the Very Large Array (VLA). 
The first uGMRT epoch (300$-$820\,MHz, April 2023) revealed an inverted radio spectrum which, combined 
with earlier (2021--2022) VLA data (1.5--11 GHz), unveiled a double-peaked spectrum potentially 
indicative of multi-epoch jet activity.
A second uGMRT campaign (August 2023), simultaneous with new VLA observations (1.5--11~GHz), instead revealed a flat low-frequency and a peaked high-frequency spectrum, ruling out this interpretation.
While limited by the two uGMRT epochs, variability analysis favors intrinsic jet processes, indicating a highly relativistic, closely aligned jet ($\theta<3$~deg, $\delta>19.3$, $\Gamma>9.7$). 
The equipartition magnetic field ($> 1$~mG) exceeds the equivalent Cosmic Microwave Background field at $z\sim7$ (0.2~mG), indicating synchrotron losses dominate.
The inferred Doppler boosting ($\delta>19.3$) implies that J0410$-$0139 is intrinsically radio-weak. As a blazar, it traces a much larger parent population of radio quasars at $z\sim7$, detectable only with deep ($\leq \mu$Jy) next-generation radio observations.
LOFAR 144\,MHz observations (April and July 2024) yielded $\sim$2\,mJy, well below the $\sim$8\,mJy predicted from uGMRT epochs, confirming 
strong variability at rest-frame $\sim$1\,GHz.
Low-frequency monitoring will be crucial for identifying high radio intensity states suitable for future 21 cm forest studies.
\end{abstract}



\section{Introduction} 
\label{sec:intro}
Blazars are Active Galactic Nuclei (AGN) whose relativistic jets are seen at a small angle to the line of sight \citep[e.g.,][]{blandford1979,urry1995}.
A blazar is commonly defined to have the viewing angle of the jet ($\theta$) less than 1/$\Gamma$, with $\Gamma$ being the bulk Lorentz factor of the jet plasma \citep[e.g.,][]{saikia2016}. 
Due to their peculiar orientation, blazars' radiation is strongly boosted, not obscured along the jet direction, and is less affected by the attenuation due to the interaction of radio photons with the Cosmic Microwave Background \citep[CMB, e.g.,][]{ghisellini2015,wu2017}.
These particular characteristics make blazars the most powerful persistent objects in the Universe, visible even at high redshifts \citep[e.g.,][]{romani2004,sbarrato2012,belladitta2020,banados2025}. 
Blazars also serve as valuable tracers for the whole family of jetted AGN \citep[e.g.,][]{volonteri2011,ghisellini2013,caccianiga2019,diana2022}, ensuring a census, free from obscuration effects, of all the supermassive black holes (SMBHs) hosted in jetted AGN at a given redshift.

High variability at all frequencies, from the radio \citep[e.g.,][]{ciaramella2004,liodakis2017,park2017,janssen2023,tripath2024}, to the optical and infrared \citep[e.g.,][]{chandra2011,isler2017,zhang2017,mao2018,anjum2020,cai2022,su2024}, X–ray \citep[e.g.,][]{pian2002,moretti2021,mundo2023,gokus2024,das2025,marcotulli2025} and $\gamma$–ray \citep[e.g.,][]{abdo2010,bhatta2020,rajput2020_2,paliya2024,lei2024} bands, on a wide range of time scales, from minutes to years \citep[e.g.,][]{chandra2011,marscher2016} is one of the main characteristics of blazars.
Quasi–simultaneous variability in different spectral bands is often (but not always) observed, suggesting that most of the emission is produced in a single region of the jet \citep[e.g.,][]{kim2020,rajput2020,acciari2021}.
Variability can play a crucial role in studying the spectral energy distribution (SED) of blazars and in computing key jet parameters, such as power, viewing angle, Doppler boosting, brightness temperature (T$_B$). 

Powerful radio sources offer a unique opportunity to probe neutral hydrogen through HI~21~cm absorption during the Epoch of Reionization 
\citep[EoR, $z\gtrsim6$; e.g.,][]{carilli2004,ciardi2015,thyagarajan2020,soltinsky2025}, particularly from $z\sim15$ to $z\sim6$, between the formation of the first ionizing sources and the bulk ionization of the intergalactic medium (IGM).
At these cosmic epochs, the 21 cm line is redshifted to the observed MHz frequencies (e.g., at $\sim$203~MHz at $z=6$ and $\sim$177~MHz at $z=7$). 
Therefore, only radio sources that are powerful enough \citep[$\gtrsim10$~mJy, e.g.,][]{carilli2004} in these bands can be used for such a study. 

Two key radio spectral features can probe the neutral IGM's structure and properties:
\textit{i)} a general $\sim$1\% absorption by the mean neutral IGM, i.e., a broad depression in the radio continuum; \textit{ii)} deeper, narrow absorption lines (with depths $\geq$ 5\% and widths of a few km\,s$^{-1}$). 
Both the width and depth of these absorption lines are important, as they provide information about the temperature, density, and velocity structure of the IGM--similar to the diagnostics obtained from the Lyman-alpha forest \citep[e.g.,][]{mcquinn2016,villasenor2022}--and can reveal the presence of mild density inhomogeneities and the onset of cosmic structure formation.
To date, there is no measurement of the cosmological 21 cm forest signal; the only observational attempt was performed by \cite{carilli2007} on two radio quasars at $z\sim5$, reporting a non-detection. 
This is not surprising, since at that redshift the Universe is already fully ionized \citep[e.g.,][and reference therein]{bosman2022}. 
Nevertheless, the underlying technique has been demonstrated at lower redshift: recent observations have detected intervening HI~21~cm absorption from foreground galaxies along the line of sight to radio quasars \citep[e.g.,][]{guha2025}, confirming that such absorption features can be measured and used to constrain the physical conditions of the gas.

At $z\geq6$, when the Universe was less than a billion years old, only two blazars are known: PSO J030947.49$+$271757.31 at $z=6.1$ (hereafter J0309$+$2717, \citealt{belladitta2020,spingola2020}) and VLASS J041009.05$-$013919.88 at $z=6.995$ (hereafter J0410$-$0139, \citealt{banados2024,banados2025}), which is the source studied in this paper.

J0410$-$0139 showed a peculiar radio SED (see Fig.~2 in \citealt{banados2025}) with a peak around 4 GHz (observed frame) and clear signs of strong ($\sim$30\%) and fast ($<$ days in the rest frame) multi-frequency variability. 
The peak in the SED makes this object a Giga-Hertz Peaked radio Source (GPS), meaning a young and compact jetted quasar \citep[e.g.,][for a review]{odea2021}.
High-$z$ blazars with a peaked radio SED are not uncommon: J0309$+$2717 shows a potential peak in the MHz regime \citep[][]{spingola2020,gloudemans2023}, as well as the blazars J0906$+$6930 at $z=5.47$ \citep[][]{coppejans2017} and J1026$+$2542 at $z=5.25$ \citep[][]{shao2020}. 

In a genuine GPS, the optically thick part of the SED (below the turnover frequency) is usually associated with synchrotron self-absorption (SSA) processes, and provides information on the mechanism at the basis of the expansion of the jet \citep[adiabatic expansion, e.g.,][]{orienti2008}. 
As the radio source adiabatically expands, the opacity decreases, the peak frequency shifts to lower frequencies, and the radio flux density increases below the peak.
However, for J0410$-$0139 the adiabatic expansion hypothesis has been ruled out by \cite{banados2025}, confirming that the change in the radio SED is typical of a Doppler boosted source. 

To better constrain the radio SED of J0410$-$0139 below the peak observed at GHz frequencies, observations in the MHz regime (observed frame) are therefore necessary. 
In this frequency range, only an upper limit of $\sim$14~mJy from the TIFR GMRT Sky Survey \citep[TGSS;][]{intema2017} at 150~MHz is available in the literature. 
Therefore, new low-frequency observations with the upgraded Giant Metrewave Radio Telescope \citep[uGMRT;][]{swarup1991,gupta2017} and the LOw Frequency ARray \citep[LOFAR;][]{vanHaarlem_2013} were acquired to study the sub-GHz SED (obs. frame) of J0410$-$0139.
Complementary observations in the GHz regime (obs. frame) obtained with the Karl G. Jansky Very Large Array (VLA) are then crucial to constrain the physical properties of the radio emission minimizing any variability issues.

Given the compactness of J0410$-$0139 even at Very Long Baseline Array resolution (\citealt{banados2025}), the source is expected to be unresolved at the requested uGMRT, LOFAR and VLA observations. 
This ensures that the different angular resolutions across frequencies do not affect the measured flux densities, as no extended emission is expected to be resolved out.

The paper is structured as follows: in Sect.~\ref{sec:observations_ugmrt}, Sect.~\ref{sec:observations_vla} and Sect.~\ref{sec:observations_lofar} we present the new uGMRT, VLA and LOFAR observations and data reduction, respectively; in Sect.~\ref{sec:results} we discuss the results from the new observations, while in Sect.~\ref{sec:varjetprop} we present an analysis of the variability, jet and magnetic field properties of the source; Sect.~\ref{sec:conclusion} summarizes our conclusions.

Throughout the manuscript we used a flat $\Lambda$ cold dark matter ($\Lambda$CDM) cosmology with H0 = 70 km s$^{-1}$ Mpc$^{-1}$, $\Omega_m$ = 0.30, and $\Omega_{\Lambda}$ = 0.70. 
Radio spectral indices are given assuming S$_{\nu} \propto \nu^{\alpha}$, consistently with the convention used by \cite{banados2025}. 
All uncertainties are reported at 1$\sigma$.

\section{uGMRT observations and data reduction}
\label{sec:observations_ugmrt}
The uGMRT observations were conducted on April 21 and August 31, 2023 (Proposal ID 44\_034, PI: Belladitta), using uGMRT band~3 (250-500 MHz) and band~4 (550-850 MHz). During each run, we observed J0410$-$0139 for a duration of $\sim$2 hours in each band (8 hours in total). Each observing run began with a 15-minute scan of a flux density scale calibrator (3C147 or 3C48), followed by the target source, and ended with another flux density scale calibrator scan.

The data were processed using the Source Peeling and Atmospheric Modelling \citep[SPAM;][]{intema2014_spam,intema2014_spamsoftware,intema2017} pipeline, which builds on the Astronomical Image Processing System \citep[AIPS;][]{greisen2003} scripts and incorporates direction-dependent calibration and imaging. Each session consisted of both narrow-band (bandwidth of 32 MHz) and wide-band observations (200 MHz for band 3 and 400 MHz for band 4). 
Following the standard reduction steps\footnote{\url{http://www.intema.nl/doku.php?id=huibintema:spam:pipeline}}, the wide-band data are split into 6 sub-bands of $\sim$33 MHz for band 3, and 50 MHz for band 4. Band 4 has two additional sub-bands (7 and 8); however, the higher frequency causes a decrease in the efficiency of the reflecting surface of the dishes and, therefore, higher noise levels. These channels are not included in the analysis. The sub-bands 1-6 are calibrated individually, using the source model obtained from the narrow-band data, which are extracted from the image using the Python Blob Detector and Source Finder \citep[PyBDSF;][]{mohan2015}. In the data from August 31, 2023, we encountered radio frequency interference (RFI) between spectral channels 140-195 of band 4 (sub-band 5). These channels were manually flagged during the calibration process. SPAM sets the flux density scale as determined by \cite{scaife2012} using the observed flux density calibrators 3C48 and 3C147. 
Finally, we used WSClean \citep{offringa2014} to combine all calibrated wide-band images. The average frequencies for our band 3 and 4 observations are 400 and 700 MHz, respectively. 

On the final dataset we performed a single Gaussian fit using the task IMFIT of the Common Astronomy Software Applications package (CASA, \citealt{mcmullin2007}) to quantify the continuum flux density and the size of J0410$-$0139. 
The measurements of total flux densities and peak brightnesses (from the mean value of the bands and from each spectral window) obtained from the best fit are listed in Table~\ref{tab:ugmrt_data}. 
The source is unresolved at the resolution of the uGMRT observations and the beam sizes for each frequency are reported in Table~\ref{tab:ugmrt_data}. 

\begin{table*}[!h]
    \centering
    \begin{tabular}{lcccccc}
    \hline\hline
       & Frequency  & Int.\ flux density & Peak surf.\ brightness & beam size & PA (E of N) &  RMS noise \\
            &  (GHz)     &  (mJy)           & (mJy~beam$^{-1}$) & (maj$''$)$\times$(min$''$) & (deg) & (mJy~beam$^{-1}$) \\
         \hline
 \multicolumn{7}{c}{April 2023} \\
\hline
 Mean band values & 0.4 & 6.12 $\pm$ 0.10 & 6.31 $\pm$ 0.06 & 8.6$\times$5.8 & 77.3 & 0.19  \\
            & 0.7 & 4.67 $\pm$ 0.09 & 4.9 $\pm$ 0.04  &  5.4$\times$2.9 & 58.8  & 0.11  \\     
         \hline
Sub-bands values   & 0.316 & 6.85 $\pm$ 0.26 & 6.87 $\pm$ 0.15 & 11.4$\times$7.5 & 76.6  & 0.33  \\
      &   0.350 & 6.45 $\pm$ 0.32 & 6.73 $\pm$ 0.18 & 9.9$\times$6.9 & 71.1  & 0.30   \\
      &   0.385 & 6.36 $\pm$ 0.28 & 6.97 $\pm$ 0.17 & 9.1$\times$6.2 & 73.1  & 0.25 \\
      &   0.416 & 6.23 $\pm$ 0.15 & 6.18 $\pm$ 0.09 & 9.1$\times$5.7 & 77.9  & 0.19 \\
      &   0.449 & 5.75 $\pm$ 0.15 & 5.94 $\pm$ 0.08 & 8.4$\times$5.3 & 76.6  & 0.15  \\
      &   0.481 & 5.56 $\pm$ 0.22 & 5.88 $\pm$ 0.13 & 7.8$\times$4.8 & 75.8  & 0.23  \\
      &   0.578 & 4.65 $\pm$ 0.13 & 4.80 $\pm$ 0.07 & 6.5$\times$3.6 & 56.4  & 0.14  \\
      &   0.625 & 4.14 $\pm$ 0.10 & 4.22 $\pm$ 0.06 &  5.8$\times$3.2 & 59.8 & 0.10 \\
      &   0.675 & 4.16 $\pm$ 0.10 & 4.27 $\pm$ 0.05 & 5.6$\times$3.0 & 60.0  & 0.10  \\
      &   0.725 & 5.10 $\pm$ 0.16 & 5.13 $\pm$ 0.09 & 5.1$\times$2.9 & 60.6  & 0.15  \\
      &   0.773 & 4.99 $\pm$ 0.20 & 5.23 $\pm$ 0.11  & 5.0$\times$2.7 & 61.1 & 0.14   \\
      &   0.824 & 5.85 $\pm$ 0.45 & 6.04 $\pm$ 0.25  & 4.3$\times$2.7 & 64.5  & 0.29 \\
     \hline\hline
 \multicolumn{7}{c}{August 2023} \\
\hline
Mean band values  & 0.4 &  7.13 $\pm$ 0.09 & 7.31 $\pm$ 0.05  & 10.7$\times$5.8 & $-$78.6 &  0.16  \\
            & 0.7 &  6.73 $\pm$ 0.04 & 6.72 $\pm$ 0.03  & 4.5$\times$3.4 & 61.8 &  0.14  \\     
         \hline
Sub-bands values      & 0.316 & 7.51 $\pm$ 0.25 & 7.28 $\pm$ 0.13  & 14.1$\times$7.8 & $-$75.2 & 0.26  \\
      &   0.350  & 6.93 $\pm$ 0.32 & 7.5 $\pm$ 0.19  & 12.3$\times$6.7 & $-$75.9 & 0.28 \\
      &   0.385  & 7.06 $\pm$ 0.26 & 7.5 $\pm$ 0.15  & 12.2$\times$6.2 & $-$75.0 & 0.26 \\
      &   0.416  & 6.74 $\pm$ 0.16 & 7.08 $\pm$ 0.09  & 11.1$\times$5.8 & $-$77.0 & 0.17  \\
      &   0.449  & 7.01 $\pm$ 0.13 & 7.23 $\pm$ 0.07  & 10.1$\times$5.3 & $-$77.4 & 0.21 \\
      &   0.481  & 7.26 $\pm$ 0.18 & 7.27 $\pm$ 0.10  & 9.6$\times$4.8 & $-$77.9 & 0.18 \\
      &   0.578  & 7.04 $\pm$ 0.08 & 7.08 $\pm$ 0.05  & 5.2$\times$4.3 & 65.2 & 0.13  \\
      &   0.625  & 6.54 $\pm$ 0.06 & 6.62 $\pm$ 0.03  & 4.9$\times$3.9 & 69.2 & 0.11  \\
      &   0.675  & 6.53 $\pm$ 0.07 & 6.32 $\pm$ 0.04  & 5.1$\times$3.4 & 69.4 & 0.12  \\
      &   0.725  & 6.47 $\pm$ 0.07 & 6.67 $\pm$ 0.04  & 4.4$\times$3.4 & 67.7 & 0.12  \\
      &   0.773  & 6.55 $\pm$ 0.09 & 6.71 $\pm$ 0.05  & 3.9$\times$3.2 & 68.0 & 0.12  \\
      &   0.824  & 6.94 $\pm$ 0.17 & 6.90 $\pm$ 0.10  & 3.7$\times$3.0 & 63.9 & 0.17  \\
     \hline
     
    \end{tabular}
    \caption{\small uGMRT radio flux densities extracted from the radio images from the April 2023 (top) and August 2023 (bottom) observations. The quoted uncertainties represent only the statistical errors from the measurements. An additional systematic uncertainty of $\sim$10\% (due to flux density scale calibration) should be considered \citep[e.g.,][]{intema2017}.}
    \label{tab:ugmrt_data}
\end{table*}

\section{VLA observations and data reduction}
\label{sec:observations_vla}

The VLA observations of J0410$-$0139 were carried out on 2023 August 30 in its most extended, A configuration (maximum baseline $\sim$ 36\,km) at L (1--2 GHz), S (2--4 GHz), C (4--8 GHz), and X (8--12 GHz) bands (proposal ID VLA/23A-240, PI: Ba{\~n}ados). The L-band frequency range was covered 
with 16 $\times$ 64\,MHz sub-bands and 1\,MHz wide spectral channels. S-band's 2\,GHz frequency span was covered with 16 $\times$ 128\,MHz sub-bands and 2\,MHz wide spectral channels. Both C- and X-bands' 4\,GHz frequency spans were covered by 32 $\times$ 128\,MHz sub-bands and 2\,MHz wide spectral channels. The correlator integration time was set to 2\,s.

In addition to the target, the source 3C\,147 was observed to calibrate the flux density scale and the bandpass, while the source J0423$-$0120 was observed as the complex gain calibrator. The single observing session, with the four frequency band noted above, was about 2.4\,hr, and the total on-target-source time at each band was about 23\,min.

Data reduction was performed in CASA, with initial calibration and data editing being performed using the VLA CASA pipeline version 2022.2.0.64. Further manual editing and phase self-calibration were performed to produce the radio images.  
For each of the C and X-bands, two images per band were made, each spanning a frequency range of 2\,GHz. 
We used the AIPS task JMFIT to perform 2-D Gaussian fitting to measure the flux densities of the source at the various frequencies. The results are reported in Table~\ref{tab:vla_data}, where the quoted uncertainties represent only the statistical errors from the measurements. An additional systematic uncertainty of $\sim$5\% (due to flux density scale calibration) should be considered\footnote{\url{https://science.nrao.edu/facilities/vla/docs/manuals/oss/performance/fdscale}}.
The source is unresolved at the resolution of the VLA observations and the beam sizes for each frequency are reported in Table~\ref{tab:vla_data}.

\begin{table*}[!h]
    \centering
    \begin{tabular}{ccccccc}
    \hline\hline
       Band & Frequency  & Int.\ flux density & Peak surf. brightness &  beam size & PA (E of N) &  RMS noise \\
            &  (GHz)     &  (mJy)     & (mJy beam$^{-1}$        & (maj$''$)$\times$(min$''$) & (deg) & (mJy~beam$^{-1}$) \\
         \hline
 \multicolumn{7}{c}{August 2023} \\
\hline
 L & 1.5 & 7.47 $\pm$  0.02 &  7.47 $\pm$  0.02 & 1.41$\times$0.91 & -14.98 & 0.023 \\
  S & 3.0 & 8.18 $\pm$  0.02 & 8.11 $\pm$ 0.01 & 0.61$\times$0.45 & -18.98 & 0.012 \\
 C & 5.0 & 8.76 $\pm$  0.01 & 8.68 $\pm$ 0.01  & 0.43$\times$0.32 & -12.10 & 0.008 \\
 C & 7.0 & 7.83 $\pm$  0.01  &  7.78 $\pm$ 0.01 & 0.30$\times$0.23 & -13.73 & 0.007 \\
 X & 9.0 & 6.66 $\pm$  0.01 &  6.65 $\pm$ 0.01 & 0.24$\times$0.18 & -8.30 & 0.010 \\
 X & 11.0 & 6.16 $\pm$  0.01 & 6.16 $\pm$ 0.01 &  0.19$\times$0.15 & -14.17 & 0.011 \\
     \hline   
    \end{tabular}
    \caption{\small VLA radio flux densities extracted from the radio images from the August 2023 observations. An additional systematic uncertainty of $\sim$5\% (due to flux density scale calibration) should be considered beside the reported statistical errors.}
    \label{tab:vla_data}
\end{table*}

\section{LOFAR observations and data reduction}
\label{sec:observations_lofar}
J0410$-$0139 was observed with the LOFAR High Band Antenna (HBA) as part of project DDT20\_001  (PI: Belladitta) with four observations in April 2024 (13th-25th) and a further two in July 2024 (6th-14th). Each target observation was limited to a duration of 2\,hrs to ensure a high observing elevation, reducing sensitivity losses due to projection effects for this low-declination target. In each observing session, a 10\,min calibrator (3C48 or 3C147) scan was performed both before and after the observations of the target field. 

All data were recorded over a bandwidth of 120-168\,MHz, flagged for interference (\citealt{offringa2012}), averaged to a time resolution of 1\,s and a frequency resolution of  12.1875\,kHz, and compressed using Dysco (\citealt{offringa2016}) before being uploaded to the LOFAR Long-Term Archive. Subsequent data processing followed the standard procedure for LOFAR HBA observations (e.g., \citealt{shimwell2022}). Each calibrator dataset was processed using the LOFAR Initial Calibration \citep[LINC;][]{deGasperin_2019} pipeline, which derives station clock offsets, XX-YY phase offsets, and amplitude calibration solutions. 

These direction-independent solutions are then applied to the corresponding target field datasets using the LINC target pipeline, which also corrects for ionospheric Faraday rotation and performs a global phase-only calibration against a TGSS sky model of the field. Finally, direction-dependent phase and amplitude calibration solutions were derived and applied in 20 directions across the field by processing all the data together in DDF-pipeline \citep[][]{Tasse2021,shimwell2022}, which makes use of DDFacet \citep{Tasse2018} and kMS \citep[][]{Smirnov2015,Tasse2014} to correct for residual ionospheric and beam-model errors. 

Final images were created for the April 2024 and July 2024 observing epochs and for the combined full dataset. 
The April and July epoch images have RMS noise values of 210\,$\mu$Jy/beam and 360\,$\mu$Jy/beam, respectively, whereas the combined dataset reaches an RMS noise level of 180\,$\mu$Jy/beam at a resolution of 9$\arcsec\times8\arcsec$. 
Following previous studies that used the same flux-density scale calibration method, we adopt a flux-density scale uncertainty of 20$\%$ (\citealt{shimwell2022}).

We used the PyBDSF software to extract the source's flux density and size.
We performed the fit on the April and July epoch images as well as on the combined image using both April and July data. 
Results are reported in Table~\ref{tab:lofar_data}. 
The flux density measurements are all consistent with each other within 1$\sigma$. 
The source is classified as compact (i.e., unresolved) by PyBDSF and also according to the definition of \cite{shimwell2019}, since S$_{int}$/S$_{peak} <$ 1.25 $\times$ 3.1(S$_{peak}$/RMS)$^{-0.53}$.

\begin{table}[!h]
    \centering
    \begin{tabular}{ccc}
    \hline\hline
    Obs.\ date &  Int.\ flux & Peak surf.\ \\
    (2024) & density & brightness \\
         & (mJy) & (mJy~beam$^{-1}$)  \\
             \hline
April & 2.87 $\pm$ 0.49  & 2.23 $\pm$ 0.23   \\
July & 2.06 $\pm$ 0.67 & 1.86 $\pm$ 0.36 \\ 
    April$+$July  &  2.68 $\pm$ 0.49 & 2.05 $\pm$ 0.23   \\
    \hline
    \end{tabular}
    \caption{LOFAR flux densities extracted from the April 2024 observation, the July 2024 observation and from the combined image (April$+$July 2024).}
    \label{tab:lofar_data}
\end{table} 

\section{Results}
\label{sec:results}
In this section, we present the results of the new uGMRT, VLA and LOFAR observations separately, followed by a joint analysis to interpret the radio emission. We also take into account the VLA data from the multi-band 2021 and 2022 observations, as previously reported in \cite{banados2025}.

\begin{figure*}[!h]
    \centering
    \includegraphics[width=\linewidth]{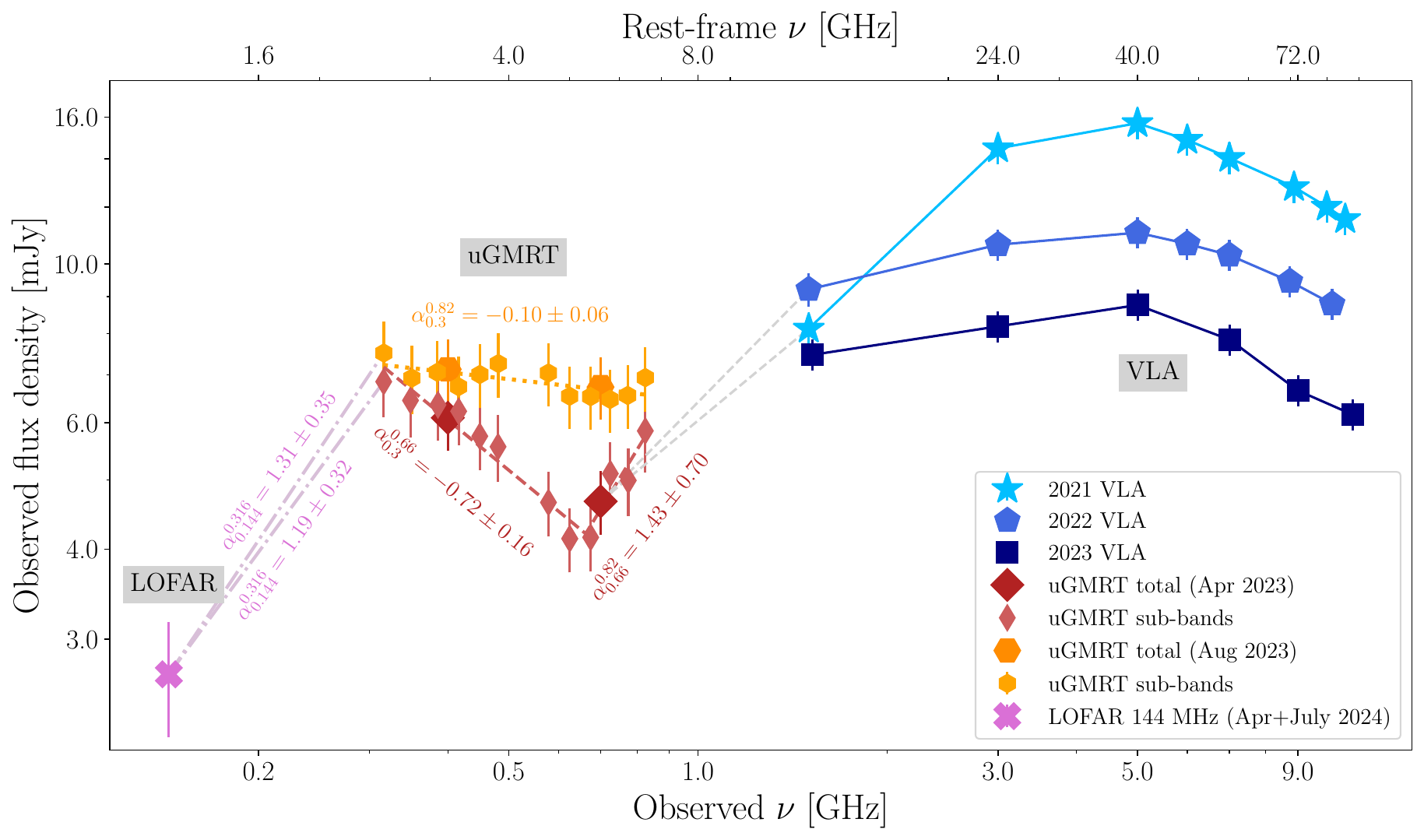}
    \caption{Radio spectral energy distribution for J0410$-$0139 from observed MHz (LOFAR and uGMRT, reported in this paper) to GHz regime (VLA from \cite{banados2025} and VLA from 2023 reported in this work). The LOFAR data point is from the combination of the observations carried out in April and July 2024; considering the large uncertainties, no flux density variability has been observed between the two observations. \textit{uGMRT}: The large diamond (hexagons) represents the overall radio flux density at 400 and 700~MHz from April 2023 (August 2023), while thin diamonds (hexagons) show the flux density in each spectral window of the two broad bands. Spectral indices computed in this paper are also shown close to the corresponding dataset. The dashed grey lines that concatenate uGMRT (April 2023) and VLA from 2021/2022 are drawn only for visual comparison. }
    \label{fig:radio_spectra}
\end{figure*}

\subsection{uGMRT -- April 2023}
\label{sec:results_ugmrt_apr2023}
The first time J0410$-$0139 was observed with the uGMRT, in April 2023 (see Sect.~\ref{sec:observations_ugmrt}), the source showed an inverted radio spectrum (i.e., a convex spectrum) as seen in Fig.~\ref{fig:radio_spectra} (dark red diamonds). 
We considered several models to fit the inverted spectrum: a simple broken power-law model, a curved power-law model \citep[e.g.,][]{callingham2017}, and a more complex model invoking SSA and free-free absorption \citep[all described in][]{shao2022}.
The best model that reproduces this inverted spectrum is a broken power-law with the following parameters: $\alpha_{0.3}^{0.66} = -0.72\pm0.16$ and $\alpha_{0.66}^{0.82} = 1.43\pm0.70$ and $\nu_{break}$ = 0.66$\pm$0.03 GHz. 

Combining the uGMRT observations with VLA data from September 2021 and/or August 2022 reported in \cite{banados2025}, results in a distinctive double-peaked radio SED for this source (see Fig.~\ref{fig:radio_spectra}): the uGMRT data reveal a convex spectrum at lower frequencies, while the VLA observations show a concave, peaked spectrum at higher frequencies.
Sources with double-peaked radio SEDs are uncommon ($<10\%$ in GPS sample, e.g., \citealt{callingham2017}) but this phenomenon is typically interpreted as signatures of multi-epoch AGN activity \citep[e.g.,][]{callingham2017,slob2022}. 
The flux density at lower frequencies is commonly attributed to past AGN activity, a relic emission from an earlier radio outburst. This component is considered non-variable, representing aged jet particles that have lost energy during their propagation far from the initial injection site. 
Conversely, the peaked SED at higher frequencies indicates recent radio activity from a newly active (young) jet expanding into the surrounding medium. This emission component is expected to be variable, as it represents the fresh population of particles recently expelled by the central SMBH.
Such sources provide crucial insights into understanding the duty cycle of AGN activity.

\subsection{uGMRT + VLA -- August 2023}
\label{sec:results_ugmrt_aug2023}
To test the interpretation mentioned in the previous section (i.e., a double-peaked radio SED as a signature of multi-epoch AGN activity), a second epoch of observations was conducted in 2023 August 31; 131 days after the initial epoch (obs.\ frame), or $\sim$16 days in the quasar rest frame. 
The goal was to determine whether the flux densities of the object remain constant across the frequency range probed by the uGMRT, confirming that the radio emission in the MHz bands originates from a relic of past radio activity. 
As shown in Fig.~\ref{fig:radio_spectra} (orange hexagons), the source not only exhibited flux density variation, but also underwent a dramatic spectral transformation: the spectrum evolved from inverted to flat ($\alpha_{0.3}^{0.82} = -0.10\pm0.06$).

This result refutes our hypothesis reported in Sect.~\ref{sec:results_ugmrt_apr2023}, establishing that the radio emission from J0410$-$0139 is variable across all observed frequencies. 
In this scenario, the convex radio SED observed in April 2023 and the flatter spectrum observed in August 2023 likely reflect the behavior of a variable jet, rather than indicating an older episode of jet activity.
Significant transitions in spectral shape are common features of the blazar population and are typically attributed to flaring episodes \citep[e.g.,][]{tinti2005,orienti2014,algaba2018}.
This motivated a detailed study of the properties of the variable emission between the two uGMRT observations, as described in Sect.~\ref{sec:varjetprop}.

\begin{figure}
    \centering
    \includegraphics[width=\linewidth]{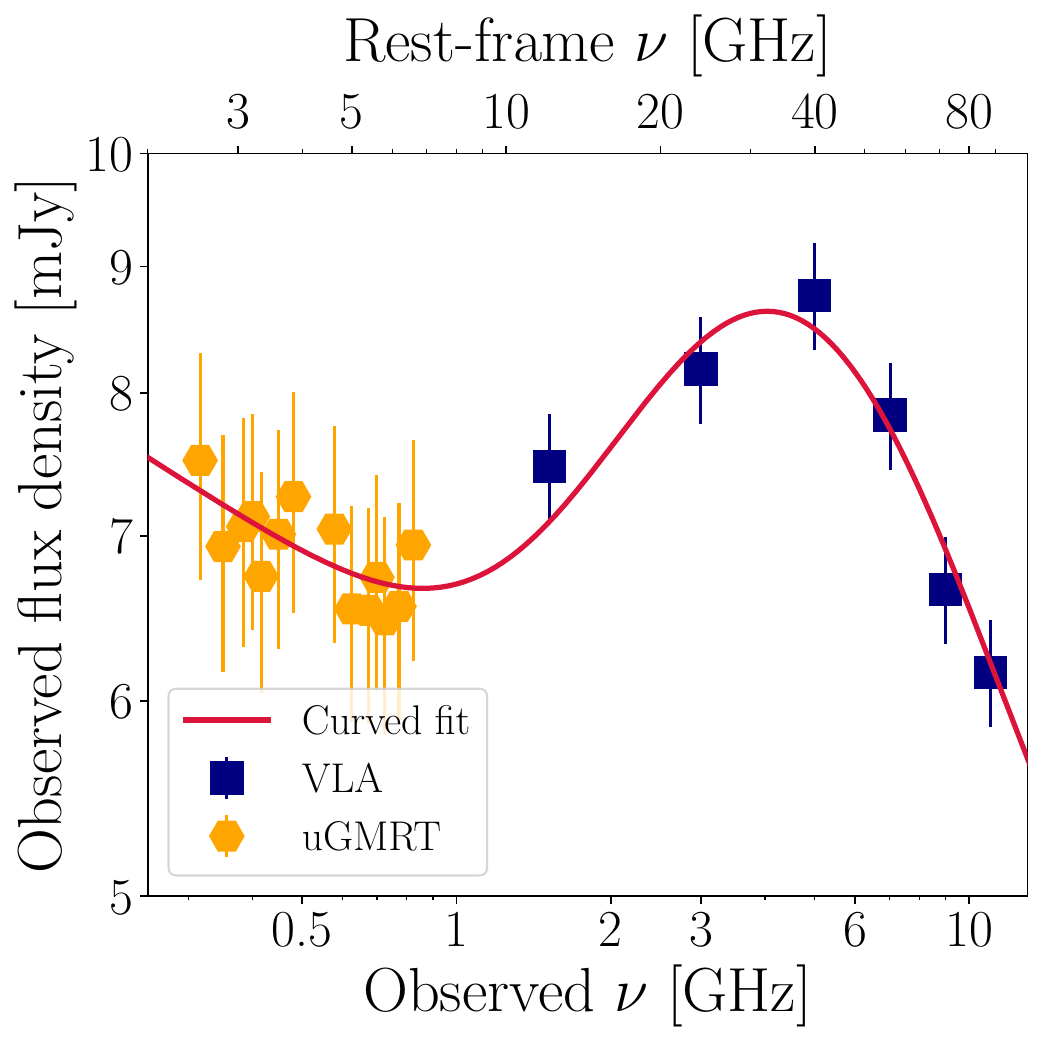}
    \caption{Quasi-simultaneous radio spectrum from the uGMRT + the VLA (2023). The radio SED shows two components: a flat power-law at lower frequencies, followed by a peaked spectrum at higher frequencies. The magenta curve represents the best fit to the data (see text and Eq.~\ref{eq:curvedspec} for details).}
    \label{fig:ugmrtvla}
\end{figure}

1.4 days (obs.\ frame) after these uGMRT observations, new VLA data were acquired.
This corresponds to just 0.17 days (4.2 hours) in the quasar rest frame.
The uGMRT+VLA spectrum (Fig.~\ref{fig:ugmrtvla}) shows two clear components: a power-law followed by a peaked spectrum.
This peaked radio SED from 1.5 to 11 GHz (obs. frame, $\sim$12-88~GHz rest-frame) is consistent with the shape obtained in 2021 and 2022 (see Fig. \ref{fig:radio_spectra} and \citealt{banados2025}), confirming, as noted by \cite{banados2025}, that the spectral evolution is incompatible with the expectations of a young radio source in adiabatic expansion \citep[e.g.,][]{orienti2020}. 

We fit this quasi-simultaneous SED from 0.3 to 11~GHz with a curved spectrum by using the equation of \cite{callingham2017}: 
\begin{equation}
S_\nu = N_{\rm c} \, \nu^{\alpha_{\rm c}} \, e^{q(\ln \nu)^2} + N_{\rm low} \, \nu^{\alpha_{\rm low}}
\label{eq:curvedspec}
\end{equation}
where $N_{\rm low}$ and $\alpha_{\rm low}$ are the normalisation and the spectral index of the power-law component, $N_{\rm c}$ is the normalisation of the curved component, $q$ is a measurement of its curvature ($|q| > 0.2$ characterizes a significantly curved spectrum; e.g., \citealt{callingham2017}) and $\alpha_{\rm c}$ is the spectral index of the curved component. The peak frequency of the peaked spectrum is given by $\nu_{\rm peak} = e^{\alpha_{\rm c}/2q}$.

The spectral fit uses a Bayesian approach with a prior on the uGMRT power-law index ($-0.10\pm0.06$, see above) and on the peak frequencies (at rest-frame frequency of $\sim$40~GHz, see \citealt{banados2025}).

We obtained the following best-fit parameters: $N_{\rm c} = 0.44\pm0.29$, $\alpha_{\rm c}=2.7\pm0.77$, $q=-0.91\pm0.26$, 
$N_{\rm low}=6.25\pm0.25$,
$\alpha_{\rm low}=-0.13\pm0.04$ and
$\nu_{\rm peak}=4.5$ GHz.
The curved spectral fit is shown in Fig.~\ref{fig:ugmrtvla}.

The value of the curvature suggests a relatively narrow electron energy distribution \citep[e.g.,][]{duffy2012}, while $\alpha_c$ indicates a very steep rising blue side of the curved component. 
The resulting $\alpha_{low}$ and $\nu_{peak}$ are consistent with our priors constraints.

\subsection{LOFAR -- April and July 2024}
Extrapolating the uGMRT radio data points toward lower frequencies, we expect a radio flux density at 144~MHz of $\sim$8~mJy (independent of the uGMRT epoch). 
Such a radio flux density is sufficient for exploratory studies of the 21 cm forest \citep[e.g.,][]{carilli2004} . 
However, the flux density measured by our LOFAR follow-up is only $\sim$2.6~mJy (see Sect.~\ref{sec:observations_lofar} and Table~\ref{tab:lofar_data}), indicating variability even at $\sim$1~GHz rest frame. 

This decrease relative to the uGMRT data points is clearly shown in Fig.~\ref{fig:radio_spectra} (pink cross). 

This would indicate a turnover in the radio SED between $\sim$0.4 and 0.144~GHz (obs. frame). 
Combined with the VLA radio SED, this LOFAR data point would constitute a second turnover. 
Multiple peaks in radio SEDs have been observed in other high-redshift sources, such as the $z=5$ blazar J1146$+$4037 (\citealt{shao2022}). 
The authors attributed the double peak to a flaring episode, consistent with the source's blazar nature. 

Given the clear variability observed at all radio frequencies to date for J0410$-$0139, the most likely explanation for the lower-than-expected LOFAR flux density remains source variability, which must occur at 144 MHz on timescales shorter than the $\sim$80 days (observed frame) or $\sim$10 days (rest frame) between our April 2024 and July 2024 observations.
Moreover, the flux density measured from the July-only data is nominally lower, but given the large uncertainties, it remains consistent with the April-only measurement.

The radio spectral index between the LOFAR and the uGMRT data points is $\alpha_{0.144}^{0.316} = 1.19\pm0.32$ or $\alpha_{0.144}^{0.316} = 1.31\pm0.35 $, considering the April 2023 or August 2023 uGMRT observations, respectively (see also Fig.~\ref{fig:radio_spectra}).

\section{Variability study and jet physical parameters}
\label{sec:varjetprop}
Leveraging the two uGMRT epochs at 400 and 700 MHz and the corresponding in-band flux density measurements, we estimate the variability of J0410$-$0139 following the methodology of \cite{ross2021}.
This enables us to place meaningful constraints on the source variability even in the absence of a well-sampled multi-frequency light curve. 
We computed the variability index parameter (VIP = 1280) and the measure of spectral shape parameter (MOSS = 242). 
The derivation of these parameters is described in Appendix \ref{app:var_radio_param}.
All estimated values exceed the thresholds for true variability established by \cite{callingham2017}.
 
Given the blazar classification of J0410$-$0139, the observed variability is most naturally attributed to intrinsic synchrotron processes in the relativistic jet. 
We can therefore use the Compton catastrophe limit on the intrinsic brightness temperature (T$_{B\_int}<10^{12}$ K; \citealt{kellermann1969}) to constrain the degree of relativistic boosting associated with the radio emission and variability of our quasar \citep[e.g.,][]{bell2019, ross2021, ighina2022}. 
The observed variability brightness temperature in the source's frame is defined by Eq.~\ref{eq:tb_final}:
\begin{equation}
T_{B,\rm var} = (1+z)\, \frac{\Delta S_\nu c^2}{2k_B\nu_{\rm obs}^2 \Omega}
\end{equation}
\noindent where $\Delta S_{\nu}$ is the variability amplitude, (1.0 and 2.1 mJy at 400 and 700 MHz, respectively),
$\nu$ is the observed frequency (400 and 700 MHz), $c$ and $k_B$ are the speed of light and Boltzmann’s constant and $\Omega$ is the solid angle subtended by the source.
The latter can be expressed as $\Omega = r^2/D_A^2$, where $r$ is the linear size of the emitting region and D$_A=1078.9$\,Mpc is the angular diameter distance to the source.

We estimate the size of the emitting region from the light-crossing time (r = $\tau$c).
The variability timescale between the two uGMRT epochs in the observed frame is 131 days, which corresponds to 16 days in the rest frame: $\Delta \tau_{rest} = \Delta \tau_{obs}/(1+z)$. 
This yields to a size $r$ of $\leq$4.15$\times$10$^{16}$cm = 0.013\,pc. 
This value is only a rough upper limit of the actual size of the varying component in the jet since it is based on the dates of the observations, which may not be representative of the intrinsic timescales of the source.

\begin{figure*}
    \centering
    {\includegraphics[width=\columnwidth]{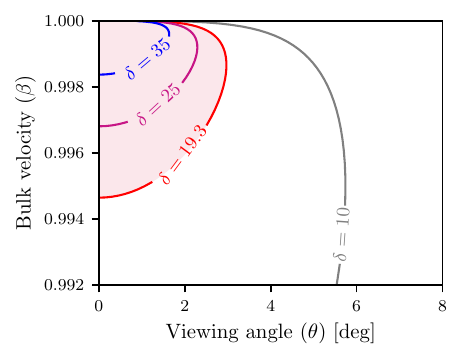}\hspace{0.1cm}
    \includegraphics[width=\columnwidth]{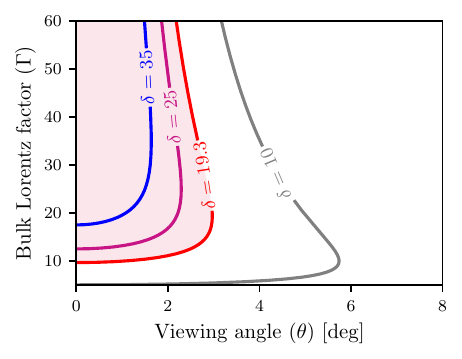}}
    \caption{Allowed values of the jet viewing angle ($\theta$) and \textit{left panel}: bulk velocity in units of the speed of light ($\beta=v/c$), \textit{right  panel}: the bulk Lorentz factor ($\Gamma$). The red shaded area corresponds to the allowed parameter space in the $\beta-\theta$ and $\Gamma-\theta$ plane for the inferred $\delta>19.3$. Curves for other values of $\delta$ are shown for comparison.}
    \label{fig:jetparam}
\end{figure*}

With these values, we derive a variability brightness temperature of $T_{B,var} \sim 7.2 \times 10^{15}$ \,K, corresponding to the most conservative estimate obtained from our measurements. This is a lower limit to the actual value since we do not know the exact dimensions of the emitting region.
Knowing the brightness temperature scales as $\delta^3T_{B\_int}$ \citep[e.g.,][]{lahteenmaki1999II,lahteenmaki1999,hovatta2009}, we estimate a lower limit on the relativistic Doppler factor of $\delta>19.3$. 
This exceeds the median values ($\delta \sim 10$) typically reported for the blazar class of Flat Spectrum Radio Quasars (FSRQ), yet remains consistent with the upper end of the distributions found in variability-selected and $\gamma$-ray-bright blazar samples \cite[e.g.,][]{rani2013,liodakis2015,liodakis2017_fgamma,liodakis2021}.
This indicates that relativistic beaming has a major role in the observed radio properties of J0410$-$0139.
From the lower limit on $\delta$ we can provide limits on the speed of the jet particles ($\beta$) and the jet’s viewing angle ($\theta$).
Figure \ref{fig:jetparam} shows the allowed $\beta-\theta$ parameter space region for $\delta>19.3$: $\beta>0.9946$ and $\theta<3$~deg.
This indicates that the jet is highly relativistic and closely aligned with our line of sight.
The lower limit on $\beta$ translates on a lower limit on $\Gamma$ (the bulk Lorentz factor of the jet) of 9.7, which is a typical value for FSRQs \citep[e.g.,][]{hovatta2009,frey2015,zhang2020,homan2021,krezinger2024}. 
All the equations describing the relationships between the different jet parameters are reported in Appendix \ref{app:var_radio_param}.

The estimated \textbf{$\delta>19.3$} corresponds to a Doppler enhancement factor ($\delta^{(3-\alpha)}$) $>9000$ (assuming $\alpha = -0.10$ from the August 2023 uGMRT observations). 
\citet{banados2025} showed that a Doppler enhancement factor of only 7.4 or 14.8 would be sufficient to classify J0410$-$0139 as intrinsically radio-weak, with its powerful radio appearance arising solely from Doppler boosting. 
Our analysis confirms this hypothesis: the blazar nature of J0410$-$0139 implies that its observed properties, such as flux density and luminosity, are not intrinsic, but highly amplified by Doppler boosting. 
The expected intrinsic flux density at 1.5\,GHz is of the order of $\lesssim 0.7\,\mu Jy$ 
depending on the observing epoch (and assuming always $\alpha = -0.10$), 
corresponding to a monochromatic luminosity $L_{1.5\,\mathrm{GHz}} \lesssim 0.4 \times 10^{30}$\,erg\,s$^{-1}$\,Hz$^{-1}$.

This value is comparable to the luminosities inferred for low-power compact radio sources in the local Universe \citep[e.g.,][and references therein]{baldi2023}, while also entering the regime where emission from compact jets overlaps with star formation \citep[e.g.,][]{panessa2019}.

Given the intrinsically weak nature of its radio emission, the case of J0410$-$0139 provides strong support to one of the scenarios derived from the \textit{blazar argument} applied in \citet{banados2025}: to reconcile the predicted number of jetted quasars with the overall number of quasars expected from the UV luminosity function is that all quasars host jets, including very weak ones that remain undetectable with current radio surveys (a hypothesis also mentioned by e.g., \citealt{sbarrato2021,wolf2024}). 
Pushing the detection threshold of radio follow-up observations to $\leq1\, \mu$Jy should therefore enable the detection of jets in essentially all quasars. 

\subsection{Equipartition magnetic field}
Following \cite{duffy2012}, we use the curvature parameter ($q$) and the peak frequency to estimate the jet's magnetic field under equipartition conditions, i.e., assuming equal energy contributions from the magnetic field and the relativistic particles \citep{pacholczyk1970}. 
Applying Eq.~\ref{eq:Heq2} and adopting the source size derived from the highest-resolution observations currently available for J0410$-$0139 (VLBA at 1.5~GHz; 14.3$\times$5.9~mas; \citealt{banados2025}), we obtain an equipartition magnetic field ($H_{eq}$) of 1.0~mG. 
Since the source remains unresolved in the VLBA observations, the major and minor axes represent upper limits on the true source dimensions, making this magnetic field estimate a lower limit.

This value is consistent with the equipartition magnetic field derived using the standard single power-law synchrotron emission formula \citep[e.g.,][]{spingola2020}, which yields B$_{eq}>0.48$~mG.

Since the energy density of the CMB scales as $\propto (1+z)^4$, inverse Compton scattering off CMB photons (IC-CMB) is expected to play a significant role at the redshift of J0410$-$0139 \citep[e.g.,][]{tavecchio2000, ghisellini2014, ighina2021_cmb,sharma2026}. 
In general, IC-CMB cooling dominates over synchrotron losses when $H_{\rm CMB} > H_{\rm eq}$. 
Using Eq.~\ref{eq:Hcmb}, we derive an equivalent magnetic field of $H_{\rm CMB}=0.2$~mG at $z=6.995$, which is significantly lower than the equipartition value. 
This suggests that energy losses within the radio jet are likely dominated by synchrotron radiation, with IC-CMB playing a negligible role.

\subsection{Interstellar scintillation}
The estimated size of the emitting region (0.013~pc) corresponds to $\sim$200 Schwarzschild radii (R$_{Sc}$)\footnote{R$_{Sc}$ = 2GM/c$^2$, where G is the gravitational constant, M is the object mass and c is the speed of light.} for the black hole mass of J0410$-$0139 ($6.9\times$10$^8 M_{\odot}$, \citealt{banados2025}). \\
This is consistent with expectations for the compact radio core in blazars (typically $\sim$100–1000 R$_{Sc}$, e.g., \citealt{hada2011,baczko2022,ricci2025}). 

Because the emitting region seems extremely compact, interstellar scintillation (ISS) may also contribute to the observed radio variability. 
ISS arises from scattering by electron density fluctuations in the turbulent ionized interstellar medium, producing flux density variations in compact radio sources \citep[e.g.,][]{rickett1990,narayan1992,walker1998,hancock2019}. 
The strength and timescale of the variability depend on the observing frequency and the scattering properties along the line of sight through the Milky Way.

We investigate the role of ISS in the observed radio variability of J0410$-$0139.
Using the NE2001 Galactic electron density model of \cite{cordes2002,cordes2003}\footnote{We made use of the python version of the code ‘pyne2001’, available at: https://pypi.org/project/pyne2001/}, we estimate a scattering measure of 0.005~kpc~m$^{-20/3}$ toward the line of sight of J0410$-$0139. 
Using a source size of 0.013~pc and an angular diameter distance of the target of 1078.9~Mpc, the predicted weak-scattering ISS timescale is only $\sim$2--3 days at 400--700 MHz, shorter than the variability implied by our observations. 
We also consider refractive ISS (RISS), which is expected to produce variability on longer timescales (days to weeks). 
For a typical scattering screen distance of 10 kpc, we obtain characteristic RISS timescales of $\sim$3--9 days.

These estimates suggest that ISS could contribute to short-term flux density fluctuations, but it is unlikely to be the dominant origin of the longer-term variability observed in J0410$-$0139. 
However, since the current uGMRT light curve comprises only two epochs, variability on timescales comparable to the ISS predictions cannot be excluded. 
Overall, while the observed flux density variation 
is consistent with intrinsic processes associated with the compact radio jet, further multi-epoch flux density monitoring is required to properly characterize the variability timescale.
 
\section{Summary and conclusion}
\label{sec:conclusion}
In this work we present radio observations from 0.144 to 11 GHz (obs.\ frame) of the most distant blazar known to date, J0410$-$0139 at $z=6.995$. 
Combining new uGMRT, LOFAR, and VLA observations obtained in 2023-2024 (presented for the first time in this work) with VLA data obtained in 2021-2022 \citep{banados2025}, we conclude that the radio SED of J0410$-$0139 is challenging to interpret.

Two uGMRT epochs (330$-$820~MHz, obs. frame) revealed variability in both flux density and spectral shape on a timescale of $\sim$16 days in the source rest frame. 
The estimated variability indices (MOSS, VIP) point to high variability, consistent with intrinsic processes within the jet plasma. 
However, because our current baseline is limited to only two epochs, a contribution from interstellar scintillation on shorter timescales cannot be excluded, as such timescales are not sampled by our observations.
Multi-epoch flux density monitoring in the future will be essential to establish the true variability timescale.

Variability analysis of the two uGMRT bands provides initial constraints on the jet parameters, revealing that the jet must be highly aligned with our line of sight ($\theta < 3$~deg) and highly relativistic ($\delta>19.3$, $\Gamma > 9.7$). These parameters are consistent with those of the blazars in the local Universe.

The variability timescale further constrains the emitting region to be extremely compact ($R < 0.013$~pc), very close to the central black hole, where the jet is still highly relativistic, collimated, and dominated by magnetic and kinetic energy \citep[e.g,][]{blandford1977,blandford1982}. 

A VLA campaign spanning 1.5--11~GHz (obs.\ frame) was conducted simultaneously with the second uGMRT epoch ($\Delta t = 0.17$~days, rest frame), revealing a flat spectrum at low frequencies followed by a peaked spectrum at high frequencies. The spectral curvature ($q = -0.91$) suggests that the energy distribution of particles injected into the jet is relatively narrow. 
Furthermore, the peak frequency ($\sim$36~GHz, rest frame) is consistent with that measured in the two previous VLA epochs (2021, 2022), once again ruling out the hypothesis of adiabatic expansion. 
This stability of the peak frequency is suggestive of a scenario where the relativistic plasma flows through a dissipation region (e.g., a recollimation shock or standing feature) where particle acceleration and magnetic field enhancement occur \citep[e.g.,][]{marscher2009,acharya2021}. 
In this framework, the radiating plasma remains relativistic and flows through this localized region, allowing for significant variability that may also be connected to high-energy flaring activity \citep[e.g.,][]{acharya2023}.

Assuming the source size inferred from VLBA observations ($14 \times 5$~mas), which represents an upper limit since the source is unresolved, and by using the parameters of the curved, quasi-simultaneous spectrum, we derive an equipartition magnetic field $H_{\rm eq} \gtrsim 0.5-1$~mG, consistent with values typically found in compact synchrotron-emitting regions in blazar jets on parsec scales. 
For magnetic field strengths $\gtrsim 0.5\text{--}1$~mG, the corresponding synchrotron cooling times at GHz frequencies are orders of magnitude longer than the observed variability timescale, implying that radiative cooling cannot be the dominant driver of the observed variability \citep[e.g.,][]{marscher2016}.  

Taken together, these results indicate that the variability is more likely driven by dynamical and geometric effects, such as changes in Doppler boosting due to relativistic motion, variations in particle injection, or the propagation of disturbances along the jet. 
The overall picture of multi-timescale and multi-frequency variability is therefore consistent with jet-based models \citep[e.g.,][]{marscher1985, marscher2014}, in which variability likely originates from compact regions near the jet base, where shocks and/or magnetic reconnection events rapidly energize particles.

The equivalent CMB magnetic field at $z = 6.995$ is estimated to be $\sim$0.2~mG, well below the equipartition value, demonstrating that IC-CMB processes do not dominate the energy losses in J0410$-$0139. 

Our analysis suggests that, already $\sim$0.8~Gyr after the Big Bang, SMBHs were capable of launching relativistic jets whose properties appear broadly consistent with those observed at lower redshift, hinting that the mechanisms responsible for jet formation operate in a largely universal way across cosmic time.

The low flux density measured by LOFAR ($\sim$2.6\,mJy), well below the $\sim$8\,mJy extrapolated from the uGMRT data, confirms that significant variability persists down to rest-frame $\sim$1\,GHz on timescales as short as $\sim$10 days. 
This variability complicates the use of J0410$-$0139 for 21 cm forest studies. 
Continued monitoring at 144\,MHz will be essential to characterize the duty cycle of high-flux states and assess the feasibility of such observations. 
As the most distant known blazar and one of the few radio sources known at $z>6$, J0410$-$0139 remains one of the most promising targets for probing the neutral intergalactic medium during the EoR with the Square Kilometre Array-LOW (SKA-LOW\footnote{\url{https://www.skao.int/en/explore/telescopes/ska-low}}).

The SKA, together with other future radio facilities, such as the next-generation VLA\footnote{\url{https://ngvla.nrao.edu/page/about}} \citep[ngVLA, e.g.,][]{selina2018,selina2021}, and the Deep Synoptic Array\footnote{\url{https://www.deepsynoptic.org/overview}} \citep[DSA, e.g.,][]{hallinan2019}, will reach sensitivity below the $\mu$Jy level, enabling the detection of low-power jets hosted by massive black holes at high redshift, as predicted by the \textit{blazar argument} \citep[e.g.,][]{spingola2026}. 
 
\section{Acknowledgments}
We thank the anonymous referee for their helpful comments and suggestions, which improved the quality of this manuscript.
This paper is based on observations with the upgraded Giant Metrewave Radio Telescope (uGMRT), under the project ID 44\_034. We thank the staff of the GMRT for their support in conducting the observations.
uGMRT is a radio facility operated by the National Centre for Radio Astrophysics (NCRA) of the Tata Institute of Fundamental Research (TIFR). 
This paper is based on observations with the LOw Frequency ARray (LOFAR), under the project ID DDT20\_001. We thank the LOFAR staff for their support in conducting the observations, and Aleksandar Shulevski in particular for his assistance.
LOFAR is the Low Frequency Array designed and constructed by ASTRON. It has observing, data processing, and data storage facilities in several countries, which are owned by various parties (each with their own funding sources), and which are collectively operated by the ILT foundation under a joint scientific policy. The ILT resources have benefitted from the following recent major funding sources: CNRS-INSU, Observatoire de Paris and Université d'Orléans, France; BMBF, MIWF-NRW, MPG, Germany; Science Foundation Ireland (SFI), Department of Business, Enterprise and Innovation (DBEI), Ireland; NWO, The Netherlands; The Science and Technology Facilities Council, UK; Ministry of Science and Higher Education, Poland; The Istituto Nazionale di Astrofisica (INAF), Italy. 
This paper is based on observations with the Karl G. Jansky Very Large Array (VLA).
The National Radio Astronomy Observatory is a facility of the U.S. National Science Foundation operated under cooperative agreement by Associated Universities, Inc.
C.M. acknowledges support from Fondecyt Iniciacion grant 11240336 and the ANID BASAL project FB210003.
C.F. acknowledges support of the Deutsche Forschungsgemeinschaft (DFG, German Research Foundation), project number 443220636, via the Research Unit FOR 5195.
This research was supported by the Munich Institute for Astro-, Particle and BioPhysics (MIAPbP) which is funded by the DFG under Germany's Excellence Strategy EXC$-$2094$-$390783311.

%

\vspace{5mm}
\facilities{uGMRT, LOFAR, VLA
}


\software{AIPS \citep{greisen2003}, astropy \citep{astropy2018},  
CASA \citep{mcmullin2007}, PyBDSF \citep{mohan2015}, SPAM \citep{intema2014_spam}
}



\appendix
\section{Derivation of physical parameters}
\label{app:var_radio_param}
The variability index parameter (VIP) is defined as:
\begin{equation}
    VIP = \left[ \sum_{i=1}^{n} \frac{(S_{1}(i) - S_{2}(i))^2}{\sigma_i^2} \right]
    \label{eq:vip}
\end{equation}
where S$_{1}(i)$ and S$_{2}(i)$ are the flux densities at epoch 1 and epoch 2, respectively, in a given sub-band $i$, and $\sigma_i$ is the combined uncertainty of each flux density added in quadrature. 
\cite{callingham2017} and \cite{ross2021} define a source to be \textit{truly variable} if VIP~$\geq 58.3$ (equivalent to a
confidence level of true variability of 5$\sigma$ in their samples of sources).

The measure of spectral shape parameter (MOSS) uses the flux density measurements directly to detect changes in the spectral shape (e.g., \citealt{ross2021}):
\begin{equation}
    MOSS = \left[ \sum_{i=1}^{n} \frac{ (\widetilde{diff}  - diff(i))^2  }{\sigma_i^2}]\right]
    \label{eq:moss}
\end{equation}
where $\widetilde{diff}$ is the median of the differences between the flux densities over all frequencies, \textit{diff(i)} is the difference of the flux densities between the two epochs at frequency $i$. 
In practice, the MOSS parameter measures the number of flux density points that are more than 1$\sigma$ away from the median difference value. 
A larger MOSS value indicates a greater spread of the difference in measurements from the median value between the two epochs, and thus a change in spectral shape. 
\cite{callingham2017} and \cite{ross2021} find that MOSS values $\geq 36.7$ (corresponding to 5$\sigma$ confidence) provide strong evidence for a clear spectral change.

In the Rayleigh-Jeans regime, the brightness temperature is, by definition: 
\begin{equation}
T_B = \frac{I_{\nu}c^2}{2k_B\nu^2} = \frac{S_{\nu}c^2}{2k_B\nu^2\Omega}  
\label{eq:tb}
\end{equation}
where I$_{\nu}$ is the specific intensity (surface brightness), equal to S$_{\nu}/\Omega$ for a source of observed solid angle $\Omega$. 
Considering the invariance of $I_\nu / \nu^3$ \citep[e.g.,][]{ghisellini2013book}, we have $I_{\nu,\text{rest}} = (1+z)^3 I_{\nu,\text{obs}}$. Substituting this transformation into Eq.~\ref{eq:tb} yields the brightness temperature in the source rest frame:
\begin{equation}
  T_{B,\rm rest} = \frac{c^2}{2k_B \nu_{\rm rest}^2} I_{\nu,\rm rest}
  = \frac{c^2}{2k_B (1+z)^2\nu_{\rm obs}^2}\,(1+z)^3 I_{\nu,\rm obs}
  = (1+z)\, \underbrace{\frac{c^2}{2k_B \nu_{\rm obs}^2} I_{\nu,\rm obs}}_{\displaystyle T_{B,\rm obs}}
\label{eq:tb2}
\end{equation}
We can now express the intensity in terms of flux density and solid angle, and the latter in terms of the physical dimension of the source ($r$) and the angular diameter distance ($D_A$): $\Omega = r^2 / D_A^2$. 
If the source shows variability, its physical dimension can be constrained using the light-crossing time argument: $r \le c \,\Delta \tau_{\text{rest}}$. 
Accounting for the flux density variability amplitude ($\Delta S_\nu$; e.g., \citealt{lahteenmaki1999II,lahteenmaki1999}), we obtain the variability brightness temperature:
\begin{equation}
    T_{B,\rm var} = (1+z)\, \frac{\Delta S_\nu c^2}{2k_B\nu_{\rm obs}^2 \Omega} = 
    (1+z)\, \frac{\Delta S_\nu D_A^2}{2k_B\nu_{\rm obs}^2 \Delta\tau_{rest}^2} = (1+z)^3\, \frac{\Delta S_\nu D_A^2}{2k_B\nu_{\rm obs}^2 \Delta\tau_{obs}^2}
\label{eq:tb_final}
\end{equation}
This can also be expressed in terms of luminosity distance [$D_L = D_A \times (1+z)^2$]:
\begin{equation}
    T_{B,\rm var} = \frac{\Delta S_\nu\, D_L^2}{2 k_B \nu_{\rm obs}^2 \,\Delta t_{\rm obs}^2\,(1+z)}, 
\end{equation}
which is equivalent to the definition of \cite{hovatta2009}.\\
The variability brightness temperature in the source's frame is then related to the Doppler boosting factor and to the intrinsic brightness temperature \citep[e.g.,][]{lahteenmaki1999II,lahteenmaki1999} by :
\begin{equation}
  \delta_{\rm var} = \left(\frac{T_{B,\rm var}}{T_{B,\rm int}}\right)^{1/3}.
\end{equation}

\noindent The Lorentz factor ($\Gamma$) is related to the bulk velocity ($\beta$) as:
\begin{equation}
    \Gamma = (1-\beta^2)^{(-1/2)}
    \label{eq:gamma}
\end{equation}
and the Doppler factor ($\delta$) is related to $\Gamma$ via:
\begin{equation}
    \delta = [\Gamma(1-\beta \cos(\theta)]^{-1}
    \label{eq:delta}.
\end{equation}

\noindent By assuming that the contributions of the magnetic field and the relativistic particles are equal \citep{pacholczyk1970}, it is possible to estimate the so-called equipartition magnetic field as: 
\begin{equation}
B_{\mathrm{eq}} = \left[ 4.5(1+\eta)c_{12}\frac{L}{V} \right]^{2/7}\; (\mathrm{G}) 
\label{eq:Heq},
\end{equation}
where $\eta=1$ (meaning that proton and electron energies are assumed to be the same), $c_{12}$ is a constant depending on spectra index and frequencies range \citep{pacholczyk1970}, 
V is the volume of the radio emitting source in cm$^3$ and $L$ is the bolometric radio luminosity (erg~s$^{-1}$) between $\nu_1=10$~MHz and $\nu_2=100$~GHz, defined as: 
\begin{equation}
L = 4\pi D_L^2 (1+z)^{-(1+\alpha)}
\int_{\nu_1}^{\nu_2} S_0
\left(\frac{\nu}{\nu_0}\right)^{\alpha} \, d\nu \,(\mathrm{erg~s^{-1}}) 
\end{equation}
where $D_L$ is the luminosity distance, $z$ is the redshift, and $S_0$ is the flux density at the observing frequency $\nu_0$.

Eq.~\ref{eq:Heq} assumed a standard synchrotron power-law for the radio emission.
When a curved spectrum is present, another way to compute the equipartition magnetic field is to use the prescription of \citep{duffy2012}: 
\begin{equation}
H_{\mathrm{equip}} =
2.01 \left( \frac{e^{1/2q}}{q} \, \nu_{\max}^{2} \, L_{\nu_{\max}} \right)^{1/7}
= 6.98
\left( \frac{e^{1/2q_{0.2}}}{q_{0.2}} \right)^{1/7}
\left( \frac{\nu_{\max}}{1\,\mathrm{GHz}} \right)^{1/7}
\left( \frac{L_{\nu_{\max}}}{10^{-35}\,\mathrm{W\,m^{-3}\,Hz^{-1}}} \right)^{2/7}
\; (\mathrm{nT})
\label{eq:Heq2}
\end{equation}
with $q_{0.2} \equiv q/0.2$, $\nu_{max}$ and $L_{\nu_{max}}$ is the emissivity at the peak frequency ($\nu_{max}$) of the radio SED.\\
The magnetic field due to the CMB can be estimated following \cite{ghisellini2014}: 
\begin{equation}
    H_{CMB} = 3.26 \times 10^{-6} \times (1+z)^2\; (\mathrm{G})
    \label{eq:Hcmb}
\end{equation}
At $z=6.995$, it yields a value of 0.2~\text{mG}.




\bibliography{main}{}
\bibliographystyle{aasjournal}



\end{document}